\documentclass[reprint,aps,prb,twocolumn,amsmath,amssymb,showpacs,preprintnumbers]{revtex4-2}

\usepackage{graphicx}[]% Include figure files
\usepackage{dcolumn}% Align table columns on decimal point
\usepackage{bm}% bold math
\usepackage{multirow}

\usepackage{braket}
\usepackage{color}
\usepackage{ulem}
\usepackage{url}
\usepackage{hyperref}

\begin{document}

\title{
Double-$Q$ and quadruple-$Q$ instabilities at low-symmetric ordering wave vectors\\ under tetragonal symmetry
}
\author{Satoru Hayami}
\affiliation{
Graduate School of Science, Hokkaido University, Sapporo 060-0810, Japan
}

\begin{abstract}
Multiple-$Q$ states are expressed as a superposition of spin density waves at multiple ordering wave vectors, which results in unconventional complicated spin textures, such as skyrmion, hedgehog, and vortex. 
We investigate the multiple-$Q$ instability by focusing on the low-symmetric ordering wave vectors in momentum space. 
By systematically performing the simulated annealing for effective spin models with various ordering wave vectors on a two-dimensional square lattice, we classify the magnetic phase diagram into four types according to the position of the ordering wave vectors. 
Three out of four cases lead to a plethora of isotropic multiple-$Q$ instabilities yielding collinear, coplanar, and noncoplanar double-$Q$ and quadruple-$Q$ magnetic phases, while the remaining case leads to an anisotropic double-$Q$ instability when the multiple-spin interaction is introduced. 
Our results indicate that exotic multiple-$Q$ phases distinct from the skyrmion crystal phase are expected when the ordering wave vectors lie on the low-symmetric positions in the Brillouin zone.

\end{abstract}

\maketitle

\section{Introduction}

Multiple-$Q$ states characterized by a superposition of multiple spin density waves have attracted much attention as a source of complicated magnetic structures. 
Depending on the type of constituent spin density waves, a variety of spin textures have been realized~\cite{gobel2021beyond}, such as a magnetic skyrmion crystal~\cite{skyrme1962unified, Bogdanov89, bocdanov1994properties, rossler2006spontaneous, Muhlbauer_2009skyrmion, yu2010real, heinze2011spontaneous, nagaosa2013topological, Tokura_doi:10.1021/acs.chemrev.0c00297}, hedgehog crystal~\cite{kanazawa2016critical, Kanazawa_PhysRevB.96.220414, fujishiro2019topological, Ishiwata_PhysRevB.101.134406, Shimizu_PhysRevB.103.054427, Aoyama_PhysRevB.103.014406}, meron crystal~\cite{Lin_PhysRevB.91.224407, yu2018transformation, Hayami_PhysRevLett.121.137202, Hayami_PhysRevB.104.094425, chen2023triple}, and tetra-axial vortex crystal~\cite{hayami2021phase}. 
For example, a two-dimensional superposition of the three spiral waves on a triangular lattice leads to the skyrmion crystal, while a three-dimensional superposition on a cubic lattice leads to the hedgehog crystal. 
These spin textures often induce topologically nontrivial physical phenomena like the topological Hall effect~\cite{Ohgushi_PhysRevB.62.R6065, Shindou_PhysRevLett.87.116801, tatara2002chirality, Neubauer_PhysRevLett.102.186602, Kanazawa_PhysRevLett.106.156603, Hamamoto_PhysRevB.92.115417, kurumaji2019skyrmion, Gobel_PhysRevB.95.094413}, which would be promising for future spintronics applications~\cite{romming2013writing, fert2013skyrmions, fert2017magnetic}. 

In a classical spin system with a fixed spin length at each site, only the isotropic exchange interactions in the form of $\bm{S}_i \cdot \bm{S}_j$, where $\bm{S}_i$ stands for the classical spin at site $i$, are not enough to stabilize the above multiple-$Q$ states in the ground state. 
In the case of the skyrmion crystal, additional factors, such as the Dzyaloshinskii-Moriya interaction~\cite{dzyaloshinsky1958thermodynamic,moriya1960anisotropic}, thermal fluctuations~\cite{Okubo_PhysRevLett.108.017206}, multiple-spin interaction~\cite{Hayami_PhysRevB.95.224424}, dipolar interaction~\cite{Utesov_PhysRevB.103.064414}, and magnetic anisotropy~\cite{leonov2015multiply, amoroso2020spontaneous, yambe2021skyrmion}, are needed to stabilize the ground-state skyrmion crystal. 
This is because the spin configurations as a result of the multiple-$Q$ superposition lead to a nonzero intensity at higher-harmonic wave vectors, which gives rise to the energy loss compared to a single-$Q$ (1$Q$) spiral state under the constraint on the spin length. 

Meanwhile, some multiple-$Q$ states do not possess the intensity at higher-harmonic wave vectors depending on the position of the ordering wave vectors in the Brillouin zone.
One of the typical examples is a triple-$Q$ state on a triangular lattice, whose ordering wave vectors lie on the Brillouin zone boundary~\cite{Martin_PhysRevLett.101.156402, Akagi_JPSJ.79.083711, Kumar_PhysRevLett.105.216405, Hayami_PhysRevB.90.060402}. 
Similar situations also occur for other lattice structures, when the ordering wave vectors lie on the Brillouin zone boundary, as found in the pyrochlore structure~\cite{Chern_PhysRevLett.105.226403}, simple cubic lattice~\cite{Alonso_PhysRevB.64.054408, Hayami_PhysRevB.89.085124}, checkerboard structure~\cite{Venderbos_PhysRevLett.109.166405}, and square lattice~\cite{Agterberg_PhysRevB.62.13816}. 
In these cases, the energies between the 1$Q$ state and multiple-$Q$ state are degenerate with each other within the bilinear isotropic exchange interaction; the degeneracy is lifted by considering infinitesimally small multiple-spin interactions~\cite{Akagi_PhysRevLett.108.096401, Hayami_PhysRevB.90.060402} and thermal fluctuations~\cite{Chern_PhysRevLett.109.156801}. 
Furthermore, another example has recently been clarified for the system with the specific low-symmetric ordering wave vectors on the square lattice, where the double-$Q$ (2$Q$) and quadruple-$Q$ (4$Q$) states without the intensity at higher-harmonic wave vectors are realized~\cite{Hayami_PhysRevB.94.024424, Hayami_PhysRevB.108.094415}. 
Although such a multiple-$Q$ instability has been shown in a specific situation, there have been no systematic investigations so far, especially for the case with the low-symmetric ordering wave vectors in the Brillouin zone. 
It is highly desired to clarify the relationship between the positions of the ordering wave vectors and the resultant multiple-$Q$ instability, which would help explore materials with complicated magnetic structures in both theory and experiments. 

In the present study, we systematically investigate the multiple-$Q$ instability on a square lattice by focusing on the low-symmetric ordering wave vectors. 
The analysis is based on the numerical simulated annealing for a simple spin model with the Heisenberg-type exchange interaction, multiple-spin interaction, and easy-axis two-spin interaction, where the multiple-spin interaction is supposed to be small. 
As a result, we classify the magnetic phase diagrams into four cases depending on the positions of the ordering wave vectors. 
We find that the ordering wave vectors $\bm{Q}=(Q^x, Q^y)$ satisfying the condition (i) $Q^x=\pi/2$ or $Q^y=\pi/2$ or (ii) $Q^x+Q^y=\pi$ results in the isotropic 2$Q$ and 4$Q$ instabilities at zero field, while the anisotropic 2$Q$ instability is caused in the case of other low-symmetric ordering wave vectors. 
We also investigate the field-induced multiple-$Q$ states in all cases. 
Our results show that a further intriguing multiple-$Q$ state is expected when ordering wave vectors lie on the low-symmetric positions in the Brillouin zone. 

The rest of this paper is organized as follows. 
In Sec.~\ref{sec: Model and method}, the spin model on the square lattice is introduced and the numerical method is outlined. 
Then, we discuss the multiple-$Q$ instability in the spin model with different low-symmetric ordering wave vectors in Sec.~\ref{sec: Results}. 
We describe the four cases according to the positions of the ordering wave vectors one by one. 
Section~\ref{sec: Summary} is devoted to a summary of the present paper.

\section{Model and method}
\label{sec: Model and method}

\begin{figure}[tb!]
\begin{center}
\includegraphics[width=1.0\hsize]{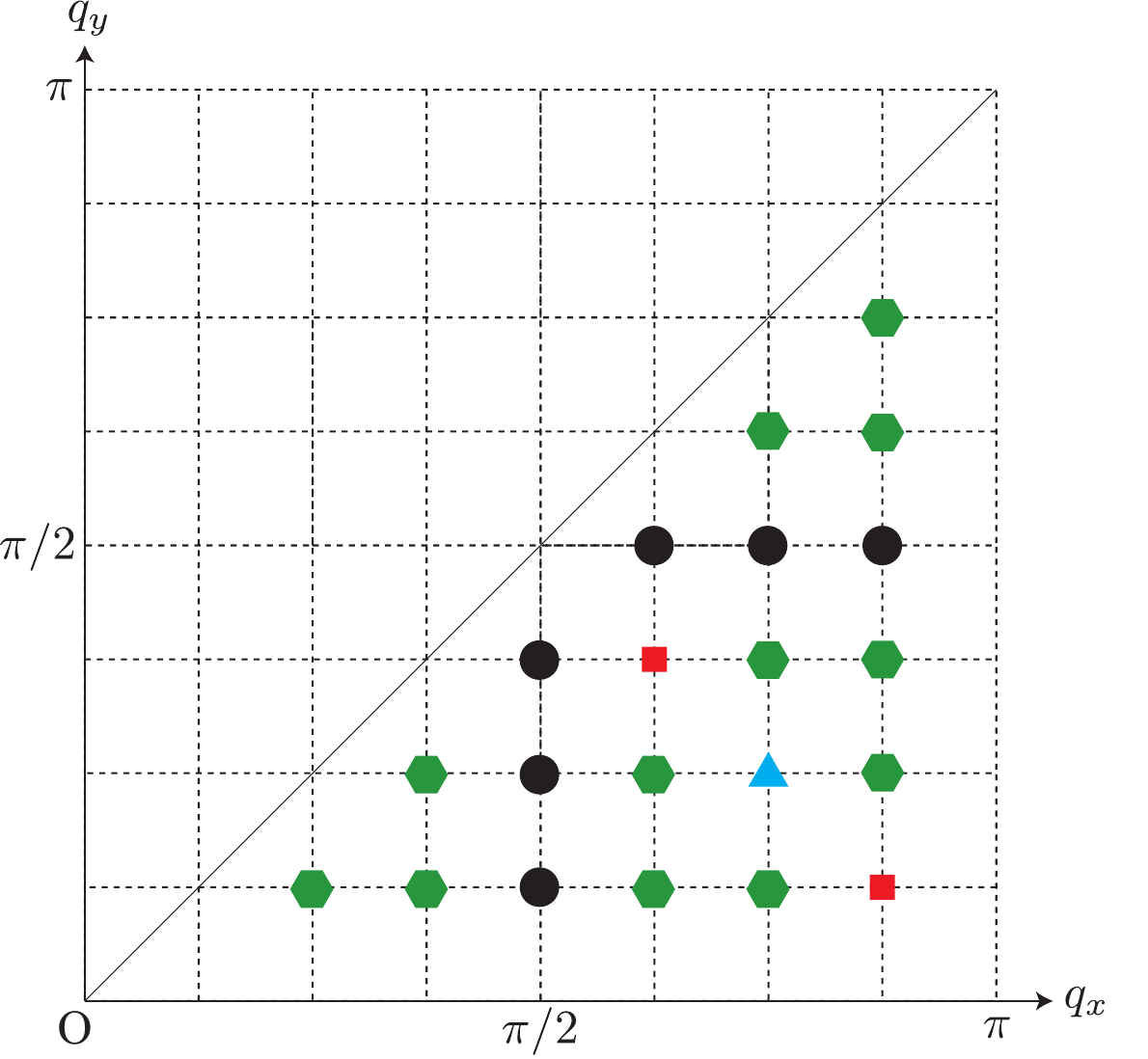} 
\caption{
\label{fig: Qvec} 
Grouping of multiple-$Q$ instability at low-symmetric ordering wave vectors under tetragonal symmetry in the first Brillouin zone. 
The green hexagons, black circles, red squares, and a blue triangle stand for Cases I, II, III, and IV, respectively. 
}
\end{center}
\end{figure}

In order to investigate the multiple-$Q$ instability at low-symmetric ordering wave vectors in a systematic way, we consider an effective spin model with the momentum-resolved interaction on the two-dimensional square lattice under the space group $P4/mmm$~\cite{hayami2021topological}, which is given by  
\begin{align}
\label{eq: Ham}
\mathcal{H}=  &-J \sum_{\nu}
( \bm{S}_{\bm{Q}_\nu}\cdot \bm{S}_{-\bm{Q}_\nu}+  I^z S^{\eta}_{\bm{Q}_\nu}S^{\eta}_{-\bm{Q}_\nu} ) \nonumber \\
&+\frac{K}{N} \sum_{\nu} ( \bm{S}_{\bm{Q}_\nu}\cdot \bm{S}_{-\bm{Q}_\nu}+  I^z S^{\eta}_{\bm{Q}_\nu}S^{\eta}_{-\bm{Q}_\nu} )^2   -  H \sum_{i}   S^z_{i}. 
\end{align}
where $\bm{S}_i=(S_i^x, S_i^y, S_i^z)$ is the localized spin at site $i$; the spin length is fixed to be $|\bm{S}_i|=1$. 
The $\bm{Q}_\nu$ component of the spin is given by $\bm{S}_{\bm{Q}_\nu}=(S^x_{\bm{Q}_\nu}, S^y_{\bm{Q}_\nu}, S^z_{\bm{Q}_\nu})$, which is obtained via the Fourier transformation to $\bm{S}_i$; $\nu$ is the index of the ordering wave vectors $\bm{Q}_\nu$. 
We take the lattice constant of the square lattice as unity. 
The first term represents the bilinear exchange interaction with the coupling constant $J$. 
We suppose that the first term originates from the Ruderman-Kittel-Kasuya-Yosida (RKKY) interaction~\cite{Ruderman, Kasuya, Yosida1957}; this term is microscopically obtained in the lowest-order perturbative expansions in terms of the Kondo coupling in the Kondo lattice model. 
We further introduce the easy-axis-type anisotropic form factor in the interaction, $I^z>0$, which arises from the relativistic spin--orbit coupling. 
On the other hand, we neglect other magnetic anisotropy, such as the bond-dependent magnetic anisotropy, for simplicity~\cite{Yambe_PhysRevB.106.174437}. 
The second term represents the biquadratic exchange interaction with the coupling constant $K$, which is obtained as the higher-order contribution of the RKKY interaction in the perturbative expansion in the Kondo lattice model~\cite{Akagi_PhysRevLett.108.096401, Hayami_PhysRevB.90.060402, Hayami_PhysRevB.95.224424}, where $N$ stands for the number of total spins in the system. 
We deal with $K$ as a perturbation term, since $K$ appears as a higher-order term than $J$ in the expansion. 
In the following, we set $J=1$ as the energy unit of the model, and $K=0$ or $K=0.02$. 
The third term in Eq.~(\ref{eq: Ham}) stands for the Zeeman coupling under an external magnetic field along the $z$ (out-of-plane) direction. 

In the model in Eq.~(\ref{eq: Ham}), we only consider the interaction at particular wave vectors that mainly contribute to the ground-state energy, which enables us to examine the multiple-$Q$ instability efficiently~\cite{Hayami_PhysRevB.95.224424}. 
We choose the ordering wave vectors located at the low-symmetric positions in the first Brillouin zone. 
Since we consider the tetragonal $P4/mmm$ symmetry, it is enough to consider the region for $q_y < q_x < \pi$ and $0<q_y< \pi$ (denoted as region I) as the independent wave vectors, as shown in Fig.~\ref{fig: Qvec}. 
We set $\bm{Q}_1$ to be the wave vector in the region I. 
From the symmetry, there are seven symmetry-related wave vectors to $\bm{Q}_1$ in the first Brillouin zone, which are defined by $\bm{Q}_2 = R(\pi/2)\bm{Q}_1$, $\bm{Q}_3=M(x)\bm{Q}_1$, $\bm{Q}_4=R(\pi/2)\bm{Q}_3$, $\bm{Q}_5=-\bm{Q}_1$, $\bm{Q}_6=-\bm{Q}_2$, $\bm{Q}_7=-\bm{Q}_3$, and $\bm{Q}_8=-\bm{Q}_4$, where $R(\pi/2)$ denotes the rotational operation by $\pi/2$ around the $z$ axis and $M(x)$ denotes the mirror operation with respect to the $xz$ plane. 

As for $\bm{Q}_1$, we consider 21 sets by supposing the system size with $N=16^2$, which are denoted as the symbols of the green hexagons, black circles, red squares, and a blue triangle in Fig.~\ref{fig: Qvec}; the different symbols mean the different multiple-$Q$ instabilities, as detailed in Sec.~\ref{sec: Results}. 
From the microscopic viewpoint, the interactions at the low-symmetric ordering wave vectors become the dominant when the nesting of the Fermi surface in itinerant electron systems occurs at the corresponding wave vectors, which have been recently found in EuNiGe$_3$~\cite{singh2023transition, matsumura2023distorted, hayami2024hybrid}. 
We ignore the effect of the interactions at other wave vectors, since their contributions to the total energy are negligible in determining the ground-state spin configuration. 
A similar phenomenological approach has been used for other models, where several mechanisms causing the multiple-$Q$ instability have been found~\cite{Hayami_PhysRevB.103.024439, Utesov_PhysRevB.103.064414, Wang_PhysRevB.103.104408, Hayami_PhysRevB.105.174437, hayami2022multiple, hayami2023widely}. 

We calculate the phase diagram while varying $I^z$ and $H$ for 21 sets of ordering wave vectors based on the simulated annealing, which minimizes the energy of the model in Eq.~(\ref{eq: Ham}) at a low temperature. 
We consider the two-dimensional square lattice consisting of $N=16^2$ spins under the periodic boundary condition. 
Starting from a random spin configuration at the temperature $T_0=1.5$, we gradually reduce the temperature with a rate $T_{n+1}=0.999999 T_{n}$ in each Monte Carlo sweep up to the final temperature $T=0.0001$, where $T_n$ is the $n$th-step temperature. 
At the final temperature, we perform $10^5$-$10^6$ Monte Carlo sweeps for measurements. 

For the obtained spin configuration, we calculate the spin structure factor, which is given by 
\begin{align}
S_s^{\eta\eta}(\bm{q})&= \frac{1}{N} \sum_{i,j} S_i^{\eta} S_j^{\eta}  e^{i \bm{q}\cdot (\bm{r}_i-\bm{r}_j)},
\end{align}
for $\eta=x,y,z$; $\bm{r}_i$ is the position vector at site $i$ and $\bm{q}$ is the wave vector in the first Brillouin zone. 
The total spin structure factor is given by $S_s(\bm{q})=S_s^{xx}(\bm{q})+S_s^{yy}(\bm{q})+S_s^{zz}(\bm{q})$. 
It is noted that $S_s^{\eta\eta}(\bm{Q}_1)=S_s^{\eta\eta}(\bm{Q}_5)$, $S_s^{\eta\eta}(\bm{Q}_2)=S_s^{\eta\eta}(\bm{Q}_6)$, $S_s^{\eta\eta}(\bm{Q}_3)=S_s^{\eta\eta}(\bm{Q}_7)$, and $S_s^{\eta\eta}(\bm{Q}_4)=S_s^{\eta\eta}(\bm{Q}_8)$. 
We also calculate the $\bm{Q}_\nu$ component of the magnetic moments from the spin structure factor, which is given by 
\begin{equation}
m^{\eta}_{\bm{Q}_\nu}=\sqrt{\frac{S^{\eta \eta}_s(\bm{Q}_\nu)}
{N}}. 
\end{equation}
In addition, we define the in-plane component of $m^{\eta}_{\bm{Q}_\nu}$ as $(m^{\perp}_{\bm{Q}_\nu})^2= (m^{x}_{\bm{Q}_\nu})^2+(m^{y}_{\bm{Q}_\nu})^2$. 
The net magnetization along the field direction is given by 
\begin{equation}
M^z=
\frac{1}{N}
\sum_i S_i^z. 
\end{equation}

\section{Results}
\label{sec: Results}

We show the multiple-$Q$ instabilities in the model with 21 sets of low-symmetric ordering wave vectors. 
By systematically performing the simulated annealing, we find that multiple-$Q$ instabilities are classified into four cases, which are denoted by the different symbols in Fig.~\ref{fig: Qvec}. 
We describe them one by one in the following subsections in Secs.~\ref{sec: Case I}--\ref{sec: Case IV}.

\subsection{Case I}
\label{sec: Case I}

\begin{figure}[tb!]
\begin{center}
\includegraphics[width=1.0\hsize]{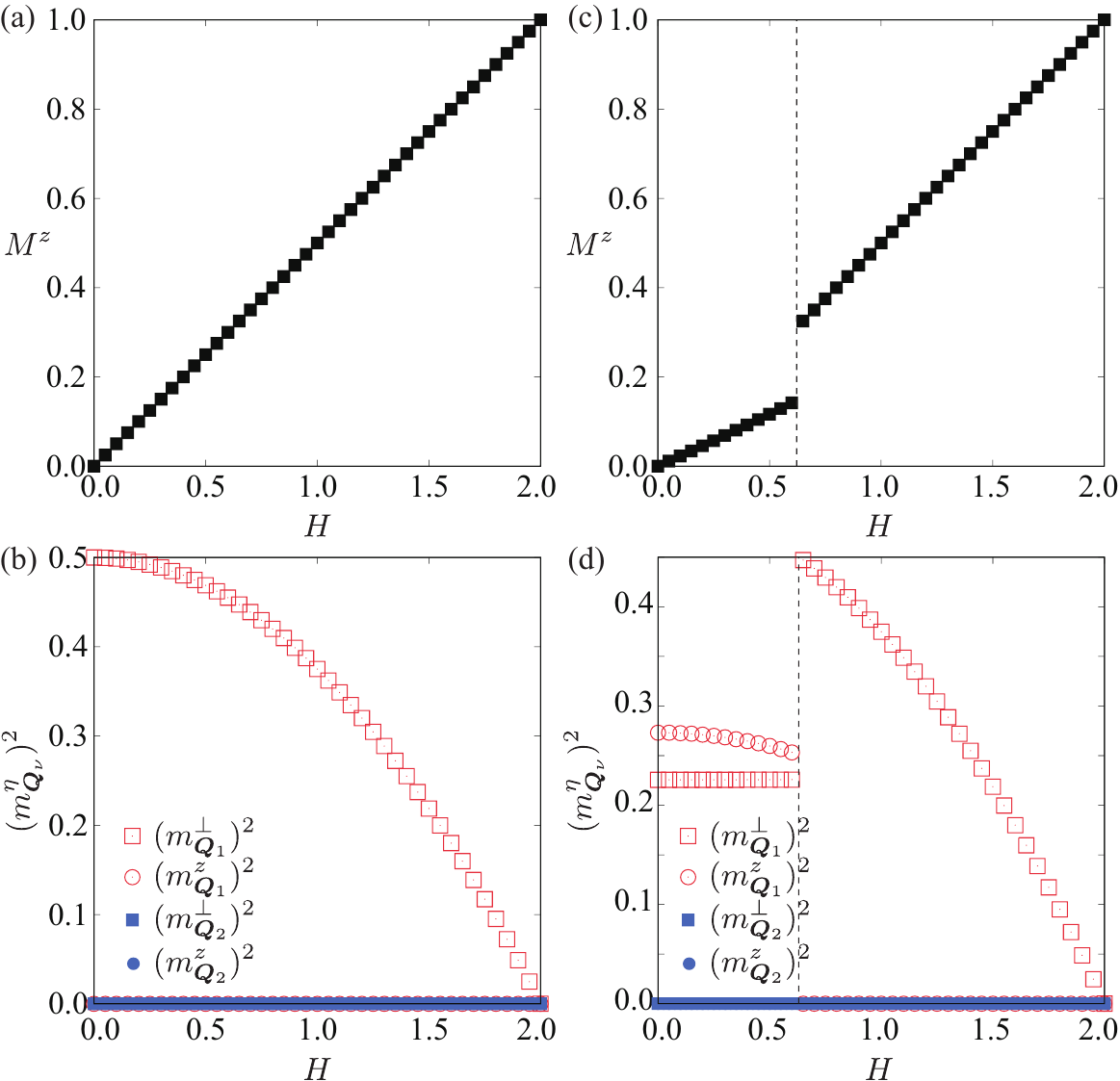} 
\caption{
\label{fig: CaseI_mag} 
$H$ dependence of (a,c) the magnetization $M^z$ and (b,d) the squared magnetic moments $(m^\eta_{\bm{Q}_\nu})^2$ for $\nu=1,2$ and $\eta=\perp, z$ in the model belonging to Case I with $\bm{Q}_1=(5\pi/8, \pi/8)$ and $\bm{Q}_2=(-\pi/8, 5\pi/8)$ at $K=0$. 
The model parameter $I^z$ is set as (a,c) $I^z=0$ and (b,d) $I^z=0.1$. 
The vertical lines represent the phase boundaries between different magnetic phases. 
It is noted that $(m^\eta_{\bm{Q}_3})^2=0$ and $(m^\eta_{\bm{Q}_4})^2=0$. 
}
\end{center}
\end{figure}

\begin{figure}[tb!]
\begin{center}
\includegraphics[width=1.0\hsize]{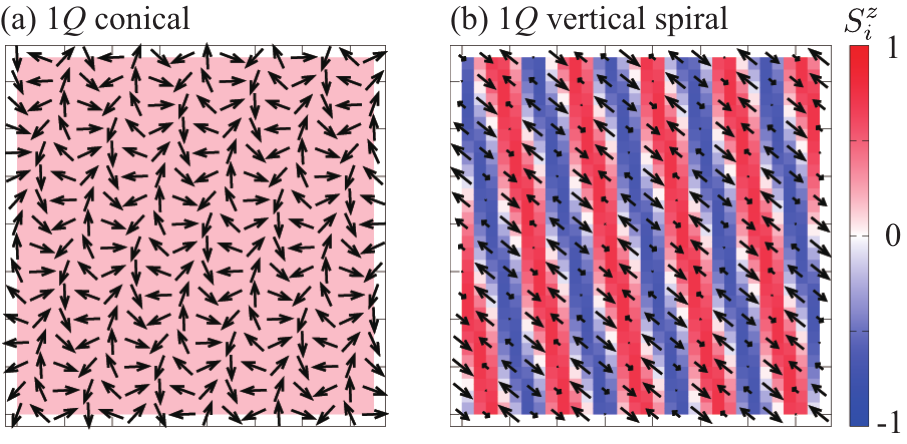} 
\caption{
\label{fig: CaseI_spin} 
Real-space spin configurations of (a) the 1$Q$ conical state at $I^z=0$ and $H=0.35$ and (b) the 1$Q$ vertical spiral state at $I^z=0.1$ and $H=0.15$. 
The arrows represent the in-plane spin components, while the color represents the out-of-plane spin component. 
}
\end{center}
\end{figure}

We consider the case where the ordering wave vectors lie at low-symmetric positions in the Brillouin zone except for $Q_1^x \neq \pi/2$, $Q_1^y \neq \pi/2$, and $Q_1^x + Q_1^y \neq \pi$, as shown by the green hexagons in Fig.~\ref{fig: Qvec}. 
For the system size with $N=16^2$, there are 12 independent ordering wave vectors. 
Among them, we show the results for $\bm{Q}_1=(5\pi/8, \pi/8)$, although different choices of the ordering wave vector for $\bm{Q}_1$ do not affect the qualitative result. 

First, we discuss the result in the absence of $I^z$ and $K$. 
Figures~\ref{fig: CaseI_mag}(a) and \ref{fig: CaseI_mag}(b) show the $H$ dependence of the field-induced magnetization $M^z$ and the squared magnetic moments $(m^\eta_{\bm{Q}_\nu})^2$ for $\nu=1,2$ and $\eta=\perp, z$, respectively. 
In Fig.~\ref{fig: CaseI_mag}(b), we show the results by appropriately sorting $(m^\eta_{\bm{Q}_\nu})^2$ for better readability. 
At zero field ($H=0$), the 1$Q$ spiral state is stabilized. 
The spin configuration is given by $\bm{S}_i =(\cos \bm{Q}_1 \cdot \bm{r}_i, \sin \bm{Q}_1 \cdot \bm{r}_i, 0)$, where the spiral plane is arbitrary owing to the spin rotational symmetry under $I^z=H=0$. 
When the effect of $H$ is introduced, the spiral plane is fixed on the $xy$ plane in order to gain the Zeeman energy. 
In other words, the spin state is characterized by the 1$Q$ conical state with $(m^{\perp}_{\bm{Q}_1})^2 \neq 0$ and $(m^{z}_{\bm{Q}_1})^2 = 0$, whose real-space spin configuration is shown in Fig.~\ref{fig: CaseI_spin}(a). 
This state remains stable up to the saturation field $H=2$, as shown in Figs.~\ref{fig: CaseI_mag}(a) and \ref{fig: CaseI_mag}(b). 
Then, this state continuously turns into the fully polarized state for $H \geq 2$. 

By taking into account the effect of $I^z$, the magnetization curve exhibits the jump at $H \sim 0.6$, as shown in the case of $I^z=0.1$ in Fig.~\ref{fig: CaseI_mag}(c), which means that an additional magnetic phase appears in the presence of $I^z$. 
In the low-field region, the behavior of $(m^\eta_{\bm{Q}_\nu})^2$ is different from that at $I^z=0$; nonzero $(m^z_{\bm{Q}_\nu})^2$ appears, which indicates that the spiral plane lies on the $xz$ or $yz$ plane corresponding to the 1$Q$ vertical spiral state. 
Indeed, one finds such a tendency in the real-space spin configuration in Fig.~\ref{fig: CaseI_spin}(b). 
The 1$Q$ vertical spiral state shows the phase transition into the 1$Q$ conical state with increasing $H$, as shown in Fig.~\ref{fig: CaseI_mag}(d). 
Thus, no multiple-$Q$ instability occurs in the absence of $K$; this is because the energy of the 1$Q$ state is lower than the multiple-$Q$ state in Case I.

\begin{figure}[tb!]
\begin{center}
\includegraphics[width=1.0\hsize]{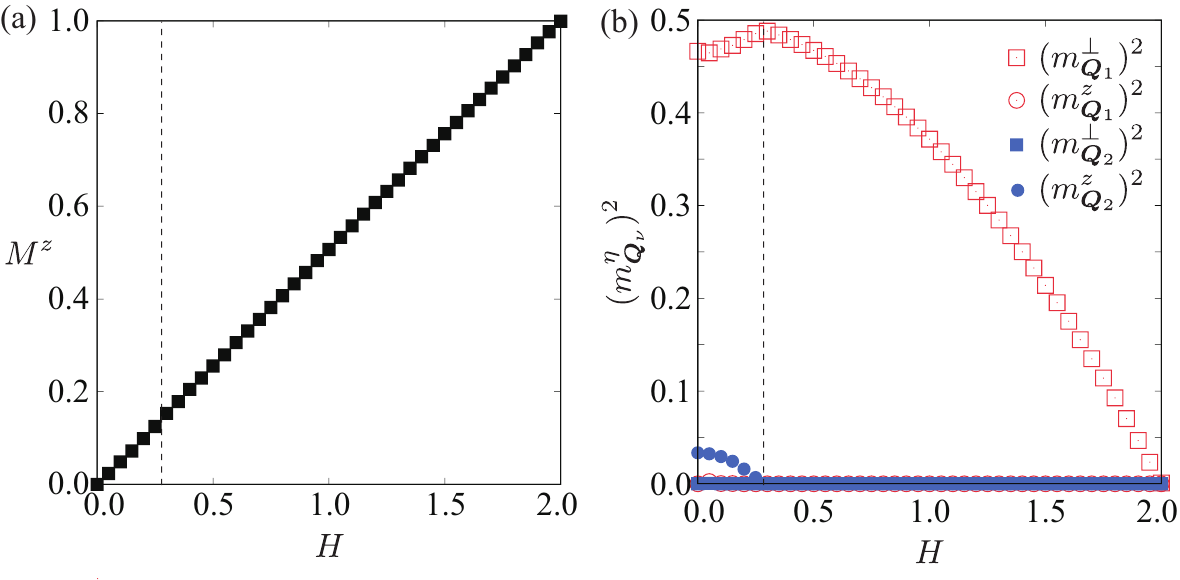} 
\caption{
\label{fig: CaseI_mag_K=0.02} 
$H$ dependence of (a) $M^z$ and (b) $(m^\eta_{\bm{Q}_\nu})^2$ for $\nu=1,2$ and $\eta=\perp, z$ at $K=0.02$ and $I^z=0$. 
The vertical lines represent the phase boundaries between different magnetic phases. 
It is noted that $(m^\eta_{\bm{Q}_3})^2=0$ and $(m^\eta_{\bm{Q}_4})^2=0$. 
}
\end{center}
\end{figure}

\begin{figure}[tb!]
\begin{center}
\includegraphics[width=1.0\hsize]{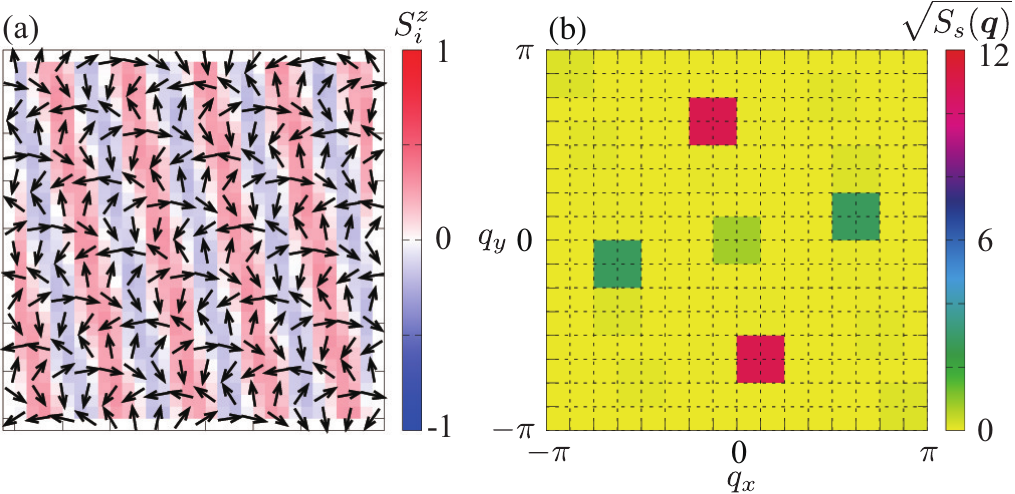} 
\caption{
\label{fig: CaseI_spin_K=0.02} 
(a) Real-space spin configurations of the anisotropic 2$Q$ state at $K=0.02$, $I^z=0$, and $H=0.1$. 
The arrows represent the in-plane spin components, while the color represents the out-of-plane spin component. 
(b) Square root of the spin structure factor corresponding to (a). 
}
\end{center}
\end{figure}

Next, we introduce the small biquadratic interaction $K=0.02$ at $I^z=0$. 
As shown in Fig.~\ref{fig: CaseI_mag_K=0.02}(a), the magnetization looks continuous against $H$, which is similar to that in Fig.~\ref{fig: CaseI_mag}(a). 
Meanwhile, $(m^z_{\bm{Q}_2})^2$ becomes nonzero in addition to $(m^{\perp}_{\bm{Q}_1})^2$ with different magnitudes, as shown in Fig.~\ref{fig: CaseI_mag_K=0.02}(b), which indicates the emergence of the anisotropic 2$Q$ state. 
The spin configuration is characterized by a superposition of the conical spiral wave in the $\bm{Q}_1$ component and the sinusoidal wave with the $z$-spin modulation in the $\bm{Q}_2$ component. 
Such a feature can be seen in the real-space spin configuration in Fig.~\ref{fig: CaseI_spin_K=0.02}(a) and the spin structure factor in Fig.~\ref{fig: CaseI_spin_K=0.02}(b), where the $\bm{Q}_2$ component of magnetic moments is dominant and the $\bm{Q}_1$ component is subdominant. 
It is noted that small intensities at higher-harmonic wave vectors, such as $\bm{Q}_2+2\bm{Q}_1$, exist owing to the superposition of the 2$Q$ spin density waves at $\bm{Q}_1$ and $\bm{Q}_2$. 
A similar anisotropic 2$Q$ state has been clarified in the case where the ordering wave vectors lie in the high-symmetric $\langle 100 \rangle $ and $\langle 110 \rangle $ lines~\cite{Ozawa_doi:10.7566/JPSJ.85.103703, Hayami_PhysRevB.95.224424}.

\subsection{Case II}
\label{sec: Case II}

\begin{figure}[tb!]
\begin{center}
\includegraphics[width=1.0\hsize]{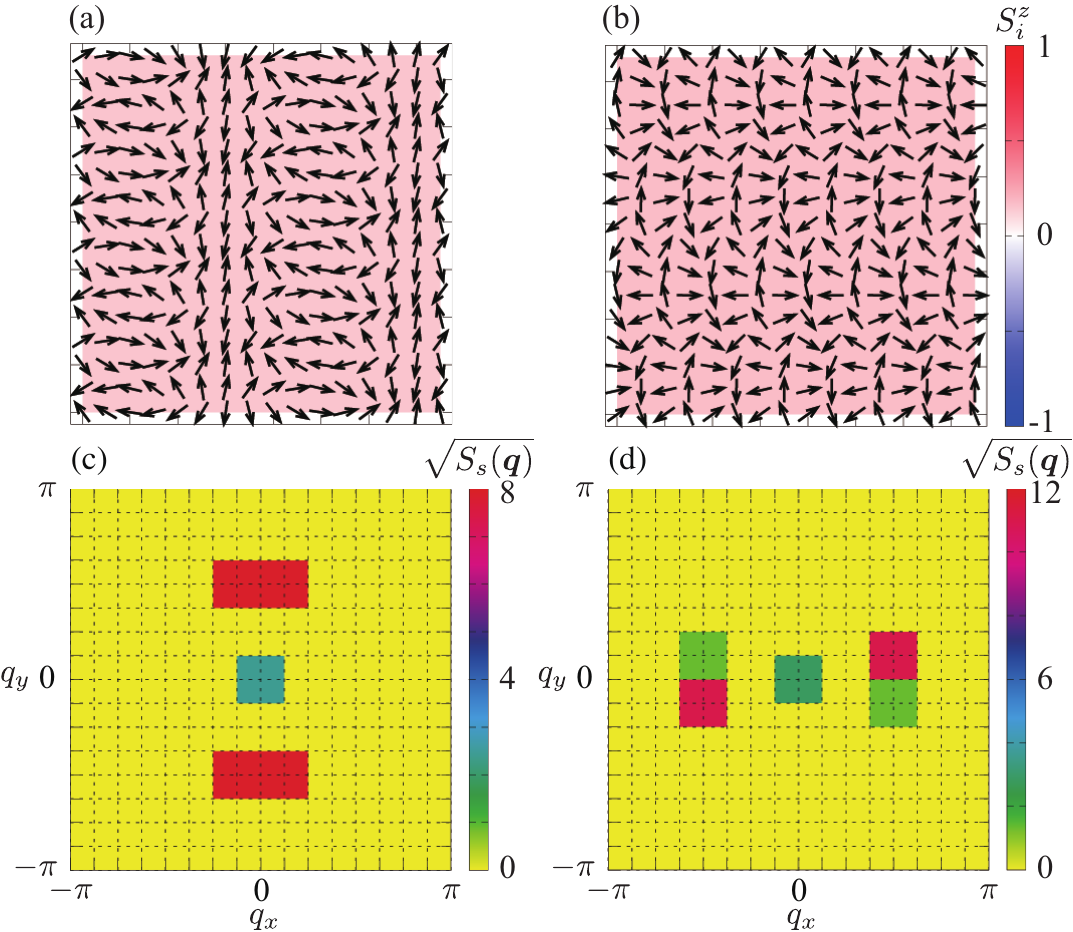} 
\caption{
\label{fig: CaseII_K=0} 
Real-space spin configurations of the 2$Q$ state at $K=0$ and $I^z=0$ for (a) $H=0.3$ and (b) $H=0.35$ in the model belonging to Case II with $\bm{Q}_1=(\pi/2, \pi/8)$. 
The arrows represent the in-plane spin components, while the color represents the out-of-plane spin component. 
(c, d) Square root of the spin structure factor corresponding to (a) and (b). 
}
\end{center}
\end{figure}

In this section, we consider the case where either $x$ or $y$ component of the ordering wave vectors takes $\pi/2$. 
There are six possibilities for $N=16^2$, as denoted by the black circles in Fig.~\ref{fig: Qvec}. 
We specifically set $\bm{Q}_1=(\pi/2, \pi/8)$, although we confirmed that similar results are obtained for other choices of $\bm{Q}_1$, as detailed below.  

In contrast to Case I, we find that a 2$Q$ state has the same energy as the 1$Q$ conical state at $I^z=0$ and $K=0$ irrespective of $H$. 
We show the real-space spin configurations and spin structure factors for nonzero $H$  in Figs.~\ref{fig: CaseII_K=0}(a)--\ref{fig: CaseII_K=0}(d), which are obtained from the simulated annealing as one of the lowest-energy states. 
As shown in Figs.~\ref{fig: CaseII_K=0}(c) and \ref{fig: CaseII_K=0}(d), this 2$Q$ state is characterized by a superposition of $\bm{Q}_1$ and $\bm{Q}_3$ [Fig.~\ref{fig: CaseII_K=0}(d)] or that of $\bm{Q}_2$ and $\bm{Q}_4$ [Fig.~\ref{fig: CaseII_K=0}(c)]. 
Thus, constituent ordering wave vectors are connected by the mirror symmetry rather than the fourfold rotational symmetry. 
We find the spin ansatz in this 2$Q$ state, which is given by 
\begin{align}
\label{eq: Si_2Q_caseII}
\bm{S}_i = \left(
\begin{array}{c}
\sqrt{1-(M^z)^2}(c_1 \cos \mathcal{Q}_1 - \sqrt{1-c_1^2} \sin \mathcal{Q}_3) \\
\sqrt{1-(M^z)^2}(-c_1 \sin \mathcal{Q}_1 + \sqrt{1-c_1^2} \cos \mathcal{Q}_3)  \\
M^z \\ 
\end{array}
\right), 
\end{align}
where $\mathcal{Q}_\nu = \bm{Q}_\nu \cdot \bm{r}_i$. 
$M^z$ represents the magnetization and $c_1$ represents the numerical coefficient satisfying $0 \leq c_1 \leq 1$. 
A similar spin ansatz is also obtained for a 2$Q$ state with nonzero $\bm{Q}_2$ and $\bm{Q}_4$ components.

The expressions in Eq.~(\ref{eq: Si_2Q_caseII}) exhibit two characteristic points. 
The first is that the spin configuration in Eq.~(\ref{eq: Si_2Q_caseII}) connects to that in the 1$Q$ conical state when setting $c_1 = 0$ or $c_1 = 1$. 
Thus, this state is regarded as a superposition of two conical spiral waves, i.e., a 2$Q$ conical state. 
The second is that the energy of this state is unchanged for $0\leq c_1 \leq 1$; the 1$Q$ and 2$Q$ conical states are energetically degenerate. 
This is also understood from the fact that the obtained 2$Q$ states do not have the intensities at higher-harmonic wave vectors like $\bm{Q}_1+\bm{Q}_3$ and $\bm{Q}_2+\bm{Q}_4$ in the spin structure factor, as shown in Figs.~\ref{fig: CaseII_K=0}(c) and \ref{fig: CaseII_K=0}(d). 
Indeed, from the expression in Eq.~(\ref{eq: Si_2Q_caseII}), one finds that the interference term arising from the normalization condition $|\bm{S}_i|=1$ is summarized as $\sin (\mathcal{Q}_1+\mathcal{Q}_3)$ and it identically vanishes. 
Such a degeneracy is not lifted by $I^z$. 
In order to lift the degeneracy between the 1$Q$ and 2$Q$ spin states, the introduction of the multiple-spin interaction $K$ is required.

\begin{figure}[tb!]
\begin{center}
\includegraphics[width=1.0\hsize]{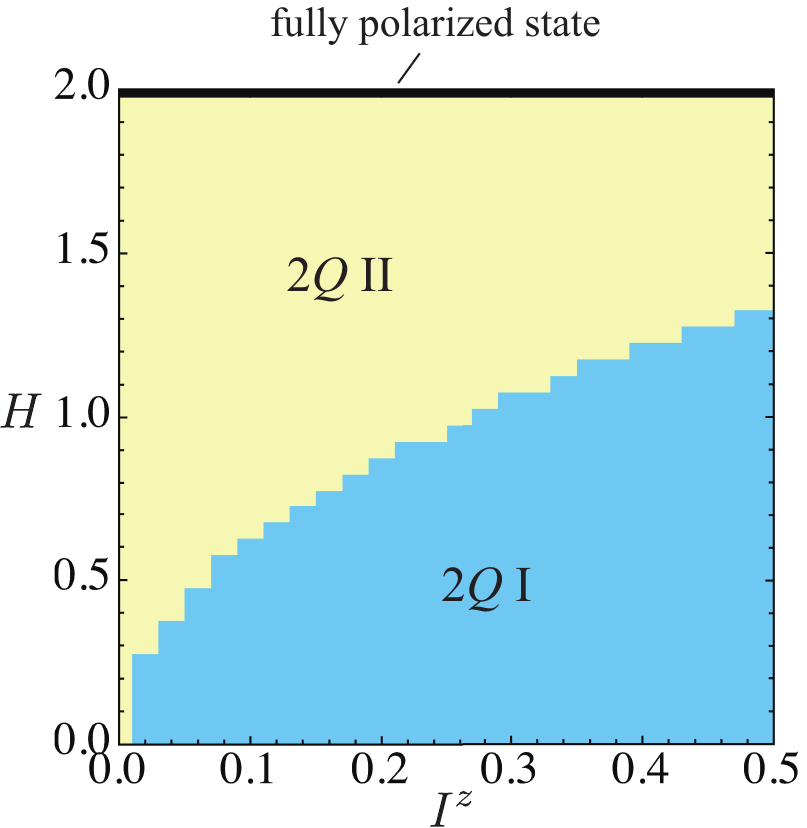} 
\caption{
\label{fig: CaseII_PD} 
Magnetic phase diagram in the plane of $I^z$ and $H$ in the model belonging to Case II with $\bm{Q}_1=(\pi/2, \pi/8)$ at $K=0.02$. 
}
\end{center}
\end{figure}

\begin{figure}[tb!]
\begin{center}
\includegraphics[width=1.0\hsize]{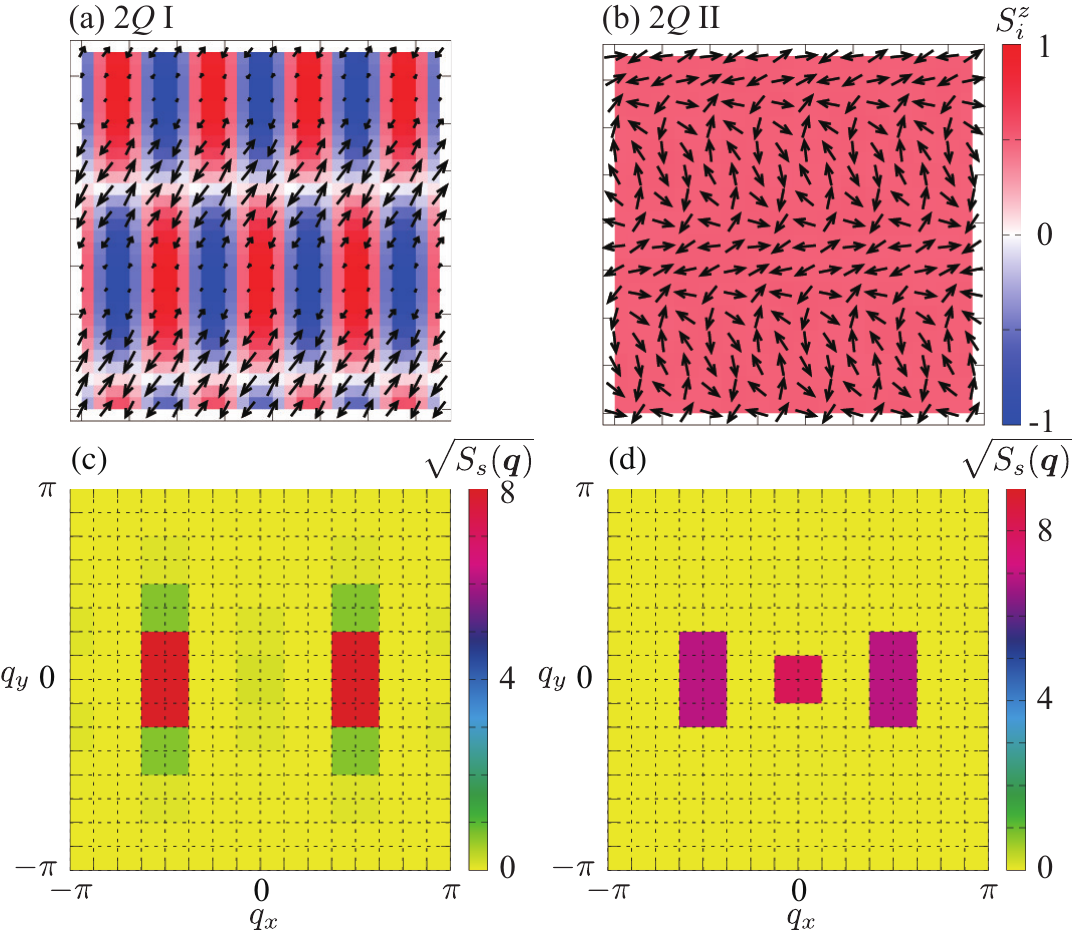} 
\caption{
\label{fig: CaseII_spin} 
Real-space spin configurations of (a) the 2$Q$ I state for $H=0.05$ and (b) the 2$Q$ II state for $H=1$ at $K=0.02$ and $I^z=0.2$. 
The arrows represent the in-plane spin components, while the color represents the out-of-plane spin component. 
(c, d) Square root of the spin structure factor corresponding to (a) and (b). 
}
\end{center}
\end{figure}

\begin{figure}[tb!]
\begin{center}
\includegraphics[width=0.93\hsize]{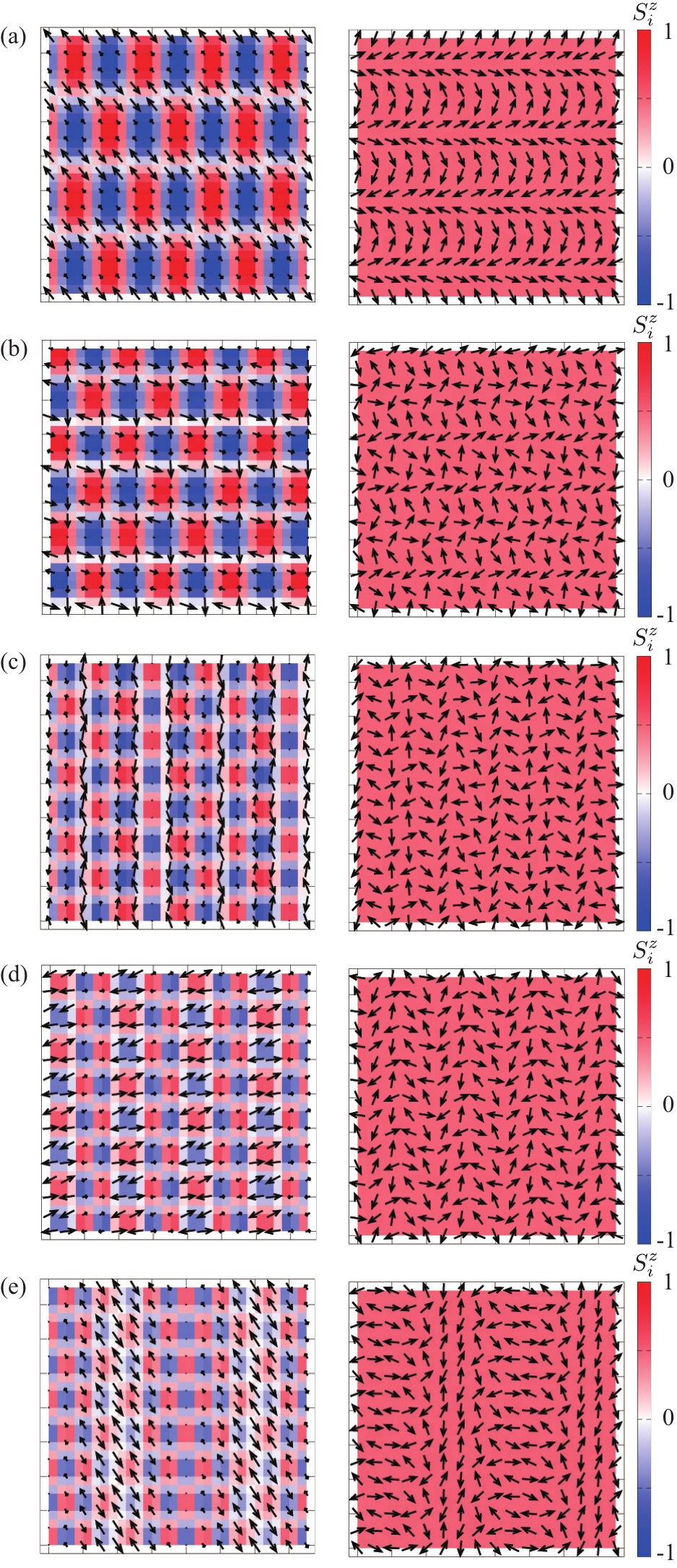} 
\caption{
\label{fig: CaseII_spin_other} 
Real-space spin configurations of (left panel) the 2$Q$ I state for $H=0.1$ and (right panel) the 2$Q$ II state for $H=1$ at $K=0.02$ and $I^z=0.2$, which are obtained in the models with different positions of the ordering wave vectors. 
The $\bm{Q}_1$ vector is given by (a) $\bm{Q}_1=(\pi/2, \pi/4)$, (b) $\bm{Q}_1=(\pi/2, 3\pi/8)$, (c) $\bm{Q}_1=(5\pi/8, \pi/2)$, (d) $\bm{Q}_1=(3\pi/4, \pi/2)$, and (e) $\bm{Q}_1=(7\pi/8, \pi/2)$. 
The arrows represent the in-plane spin components, while the color represents the out-of-plane spin component. 
}
\end{center}
\end{figure}

By introducing $K$, the energy of the 2$Q$ state becomes lower than that of the 1$Q$ state, since $K$ tends to enhance the instability toward the multiple-$Q$ states compared to the 1$Q$ state~\cite{Akagi_PhysRevLett.108.096401, Hayami_PhysRevB.90.060402, Hayami_PhysRevB.95.224424}. 
For small $K=0.02$, we construct the magnetic phase diagram against $I^z$ and $H$ in Fig.~\ref{fig: CaseII_PD}. 
There are two phases below the saturation magnetic field at $H=2$: One is the 2$Q$ I in the low-field region and the other is the 2$Q$ II state in the high-field region. 
In contrast to Case I, no single-$Q$ state appears in the phase diagram. 

In the low-field region, the 2$Q$ I state appears, whose stability region becomes wider for larger $I^z$. 
The real-space spin configuration in this state is shown in Fig.~\ref{fig: CaseII_spin}(a), where both $xy$ and $z$ spins exhibit the 2$Q$ modulation. 
Such a 2$Q$ feature is also found in the spin structure factor in Fig.~\ref{fig: CaseII_spin}(c), where the relations of $(m^{\perp}_{\bm{Q}_1})^2=(m^{\perp}_{\bm{Q}_3})^2$ and $(m^{z}_{\bm{Q}_1})^2=(m^{z}_{\bm{Q}_3})^2$ are satisfied. 
In contrast to the case at $K=0$ in Figs.~\ref{fig: CaseII_K=0}(c) and \ref{fig: CaseII_K=0}(d), there are intensities at the high-harmonic wave vectors of $\bm{Q}_\nu$; sub-dominant peak structures are found at $3\bm{Q}_1$ and $3\bm{Q}_3$, which are attributed to the fact that the additional out-of-plane modulations of both $\bm{Q}_1$ and $\bm{Q}_3$ have the different intensities from the in-plane modulations owing to $I^z$ and $H$, which results in the appearance of the elliptic spiral plane to have the $3\bm{Q}_\nu$ component.

When $H$ increases, the 2$Q$ I state turns into the 2$Q$ II state with a jump of the magnetization similar to the behavior in Fig.~\ref{fig: CaseI_mag}(c). 
The spin configuration in this state corresponds to that in Eq.~(\ref{eq: Si_2Q_caseII}) with setting $c_1=1/\sqrt{2}$; there is no $z$-spin modulation in contrast to the 2$Q$ I state. 
The real-space spin configuration and the spin structure factor are shown in Figs.~\ref{fig: CaseII_spin}(b) and \ref{fig: CaseII_spin}(d), respectively. 
This state continuously changes into the fully polarized state. 

It is noted that the emergence of the 2$Q$ state in Case II is qualitatively different from that in Case I, since, in the absence of $K$, the energy of the 2$Q$ state is degenerate with that of the 1$Q$ state in the former, while their energies are different in the latter. 
This difference is attributed from the fact that the latter 2$Q$ spin structure possesses the intensity at high-harmonic wave vectors in the spin structure factor, while the former does not. 
In this sense, the instability toward the 2$Q$ state is prominent for Case II. 
Indeed, only the 2$Q$ state appears in the phase diagram even for small $K$, as shown in Fig.~\ref{fig: CaseII_PD}. 
On the other hand, the 1$Q$ state remains stable for finite $H$ in Case I, as shown in Fig.~\ref{fig: CaseI_mag_K=0.02}(b).

The 2$Q$ I and 2$Q$ II states at $\bm{Q}_1=(\pi/2, \pi/8)$ also appear for other choices of the ordering wave vectors once either $Q^x_1$ or $Q^y_1$ takes $\pi/2$. 
We show the snapshots of the real-space spin configurations of the 2$Q$ I and 2$Q$ II states in the cases of $\bm{Q}_1=(\pi/2, \pi/4)$ in Fig.~\ref{fig: CaseII_spin_other}(a), $\bm{Q}_1=(\pi/2, 3\pi/8)$ in Fig.~\ref{fig: CaseII_spin_other}(b), $\bm{Q}_1=(5\pi/8, \pi/2)$ in Fig.~\ref{fig: CaseII_spin_other}(c), $\bm{Q}_1=(3\pi/4, \pi/2)$ in Fig.~\ref{fig: CaseII_spin_other}(d), and $\bm{Q}_1=(7\pi/8, \pi/2)$ in Fig.~\ref{fig: CaseII_spin_other}(e), where the left (right) panel shows the spin configuration of the 2$Q$ I (2$Q$ II) state. 
The spin structure factors in these states are qualitatively similar to those in Figs.~\ref{fig: CaseII_spin}(c) and \ref{fig: CaseII_spin}(d). 
Thus, the instability toward the 2$Q$ spin configurations is common when the ordering wave vectors include the $\pi/2$ modulation.

\subsection{Case III}
\label{sec: Case III}

\begin{figure}[tb!]
\begin{center}
\includegraphics[width=1.0\hsize]{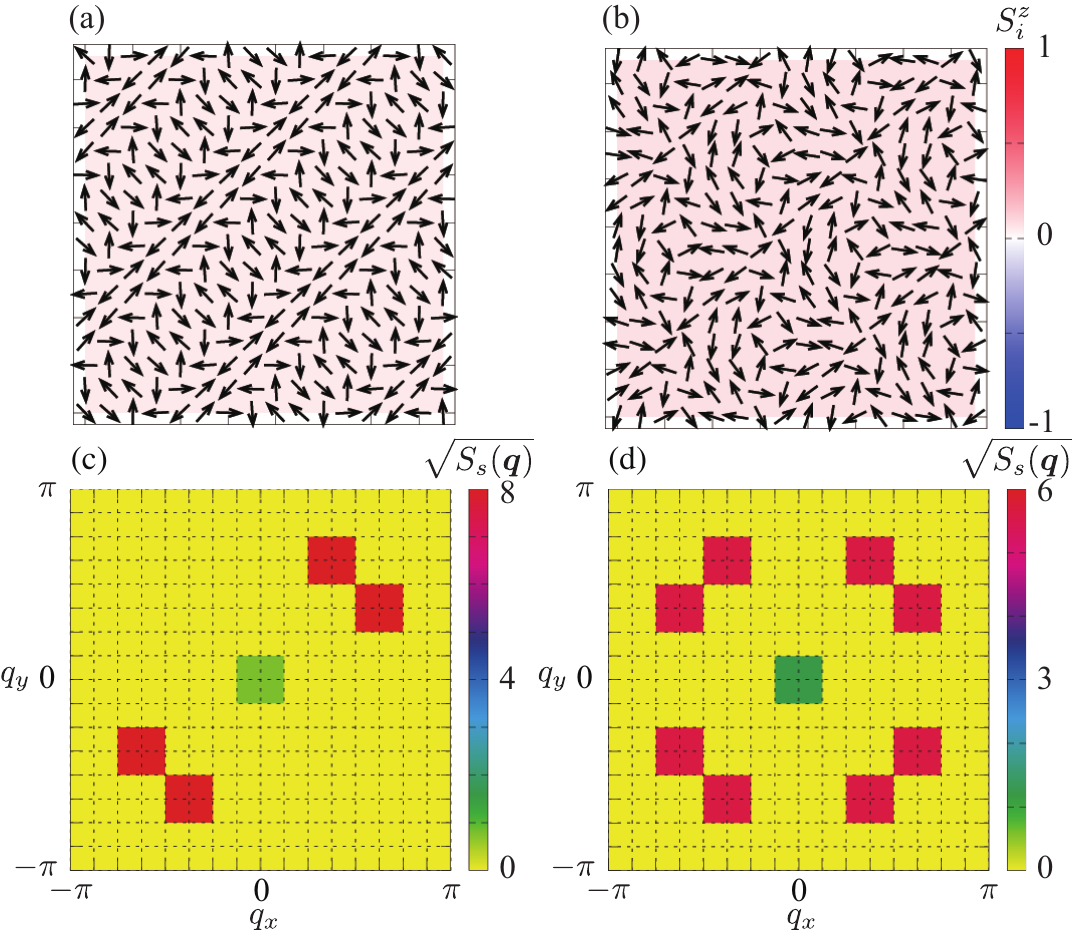} 
\caption{
\label{fig: CaseIII_K=0} 
Real-space spin configurations of (a) the 2$Q$ state for $H=0.1$ and (b) the 4$Q$ state for $H=0.15$ at $K=0$ and $I^z=0$ in the model belonging to Case III with $\bm{Q}_1=(5\pi/8, 3\pi/8)$. 
The arrows represent the in-plane spin components, while the color represents the out-of-plane spin component. 
(c, d) Square root of the spin structure factor corresponding to (a) and (b). 
}
\end{center}
\end{figure}

In this section, we discuss the situation where the ordering wave vectors satisfy $Q^x_1+Q^y_1=\pi$ except for $\bm{Q}_1=(3\pi/4, \pi/4)$, as denoted by the red squares in Fig.~\ref{fig: Qvec}; we denote this case as Case III. 
We specifically set $\bm{Q}_1=(5\pi/8, 3\pi/8)$ without loss of generality. 

For $K=0$ and $I^z=0$, we find that the 1$Q$, 2$Q$, and 4$Q$ states are energetically degenerate. 
The 2$Q$ state is represented by a superposition of the conical spiral waves at $\bm{Q}_1$ and $\bm{Q}_4$, whose spin ansatz is given by 
\begin{align}
\label{eq: 2Q_caseIII}
\bm{S}_i = \left(
\begin{array}{c}
\sqrt{1-(M^z)^2}(c_1 \cos \mathcal{Q}_1 - \sqrt{1-c_1^2} \sin \mathcal{Q}_4) \\
\sqrt{1-(M^z)^2}(-c_1 \sin \mathcal{Q}_1 + \sqrt{1-c_1^2} \cos \mathcal{Q}_4)  \\
M^z \\ 
\end{array}
\right). 
\end{align}
Although the expression is similar to that in Eq.~(\ref{eq: Si_2Q_caseII}), the constituent ordering wave vectors are different from each other; the ordering wave vectors in the present 2$Q$ state are connected by the vertical mirror plane on the $[110]$ line, while those in the 2$Q$ state in Eq.~(\ref{eq: Si_2Q_caseII}) are connected by the vertical mirror plane on the $[100]$ line. 
Meanwhile, this 2$Q$ state does not possess the intensity at high-harmonic wave vectors in the spin structure factor similar to the situation in Sec.~\ref{sec: Case II}. 
Similar to Case II, $\bm{S}_i$ is normalized as $|\bm{S}_i|=1$ for all $i$ without the normalization constant owing to the relation $\sin (\mathcal{Q}_1 +  \mathcal{Q}_4)=0$. 
The real-space spin configuration and spin structure factor are shown in Figs.~\ref{fig: CaseIII_K=0}(a) and \ref{fig: CaseIII_K=0}(c), respectively. 

The spin ansatz of the 4$Q$ state is given by 
\begin{align}
\label{eq: 4QcaseIII}
\bm{S}_i = \left(
\begin{array}{c}
\displaystyle A( \cos \mathcal{Q}_1- \sin \mathcal{Q}_2- \sin \mathcal{Q}_3 - \cos \mathcal{Q}_4) \\
\displaystyle A( \sin \mathcal{Q}_1+ \cos \mathcal{Q}_2+ \cos \mathcal{Q}_3 + \sin \mathcal{Q}_4)  \\
M^z \\ 
\end{array}
\right), 
\end{align}
where $A=\sqrt{1-(M^z)^2}/2$. 
In contrast to the 2$Q$ state in Eq.~(\ref{eq: 2Q_caseIII}), the intensities at $\bm{Q}_1$--$\bm{Q}_4$ are the same as each other, which indicates the isotropic 4$Q$ state; the isotropic intensity of the spin structure factor is found in Fig.~\ref{fig: CaseIII_K=0}(d). 
It is noted that there in no intensity at higher-harmonic wave vectors, which is understood from the expression in Eq.~(\ref{eq: 4QcaseIII}); the factor of $\cos (\mathcal{Q}_2-\mathcal{Q}_3) -\cos (\mathcal{Q}_1+\mathcal{Q}_4)+\sin (\mathcal{Q}_1-\mathcal{Q}_2)+\sin (\mathcal{Q}_1-\mathcal{Q}_3)+\sin (\mathcal{Q}_2+\mathcal{Q}_4)+\sin (\mathcal{Q}_3+\mathcal{Q}_4)$ appearing from the normalization condition of the spin length is expressed as $ c_2 \sin \pi x_i \sin\pi y_i + c_3 (\sin \pi x_i + \sin \pi y_i )$ for $\bm{r}_i=(x_i, y_i)$, which vanishes for all $i$ ($c_2$ and $c_3$ are coefficients). 
Since the modulations of $\bm{Q}_1$--$\bm{Q}_4$ occur in the $xy$ plane, this state is regarded as a 4$Q$ conical state, whose spin configuration is shown in Fig.~\ref{fig: CaseIII_K=0}(b). 
Similarly to Case II, the introduction of $K$ lifts the degeneracy among the 1$Q$ and multiple-$Q$ states.

\begin{figure}[tb!]
\begin{center}
\includegraphics[width=1.0\hsize]{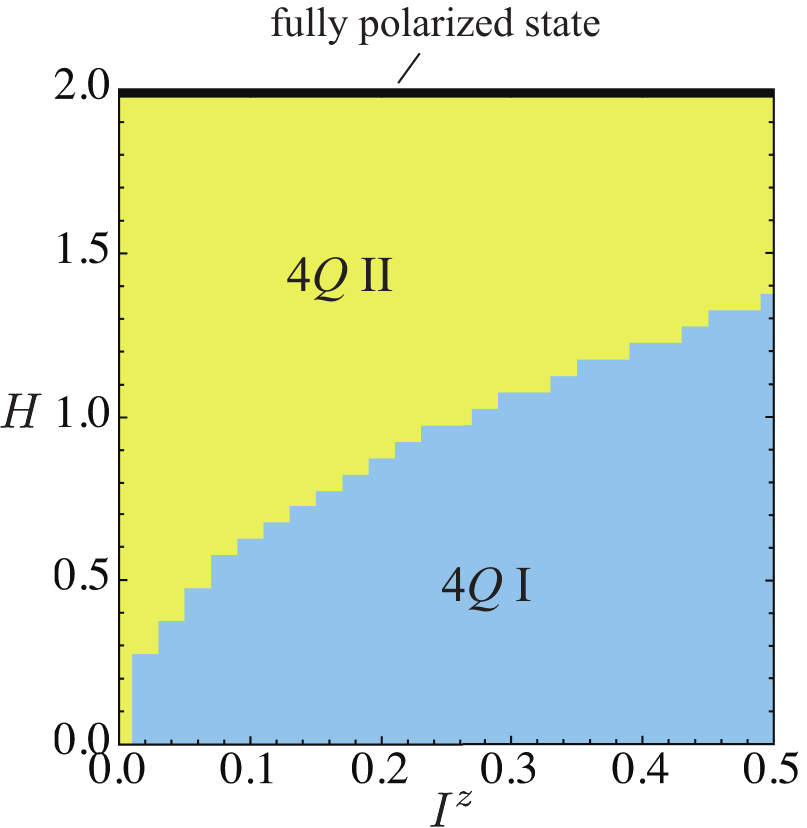} 
\caption{
\label{fig: CaseIII_PD} 
Magnetic phase diagram in the plane of $I^z$ and $H$ in the model belonging to Case III with $\bm{Q}_1=(5\pi/8, 3\pi/8)$ at $K=0.02$. 
}
\end{center}
\end{figure}

\begin{figure}[tb!]
\begin{center}
\includegraphics[width=1.0\hsize]{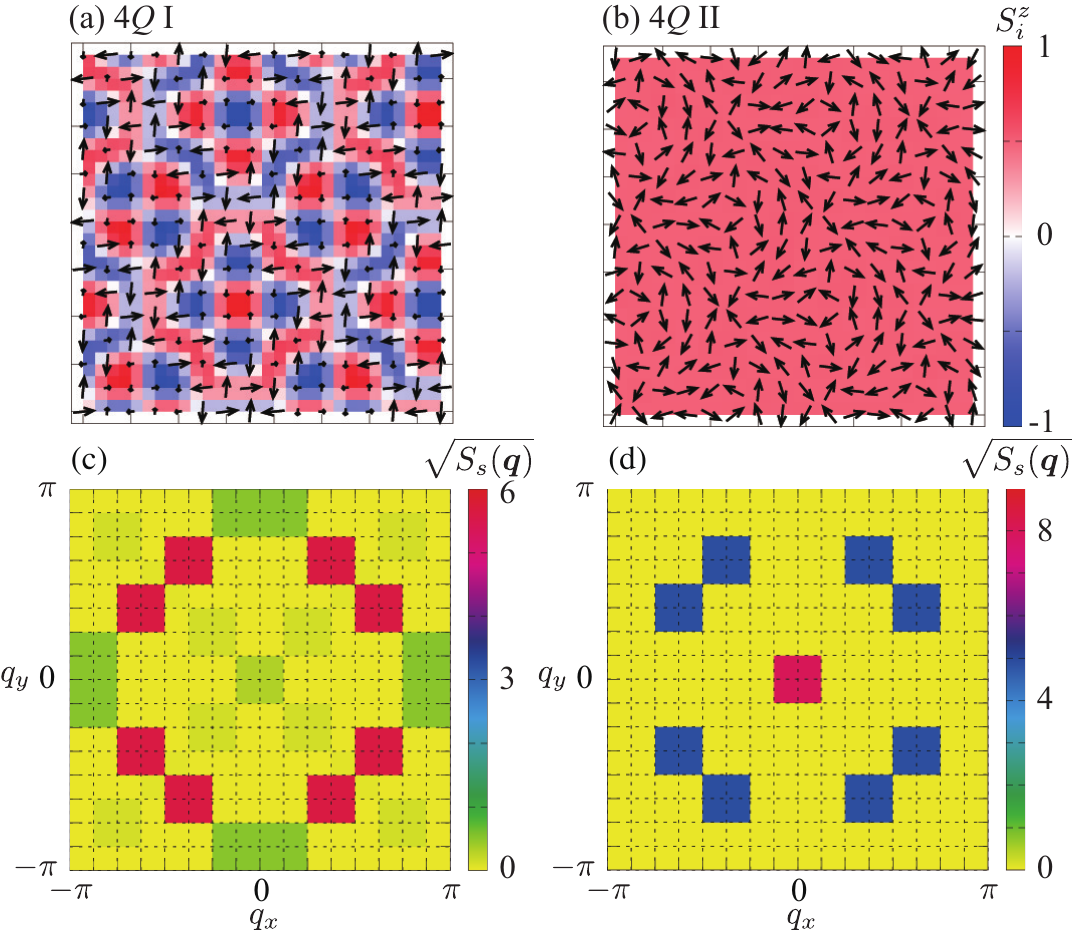} 
\caption{
\label{fig: CaseIII_spin} 
Real-space spin configurations of (a) the 4$Q$ I state for $H=0.1$ and (b) the 4$Q$ II state for $H=1$ at $K=0.02$ and $I^z=0.2$. 
The arrows represent the in-plane spin components, while the color represents the out-of-plane spin component. 
(c, d) Square root of the spin structure factor corresponding to (a) and (b). 
}
\end{center}
\end{figure}

\begin{figure}[tb!]
\begin{center}
\includegraphics[width=1.0\hsize]{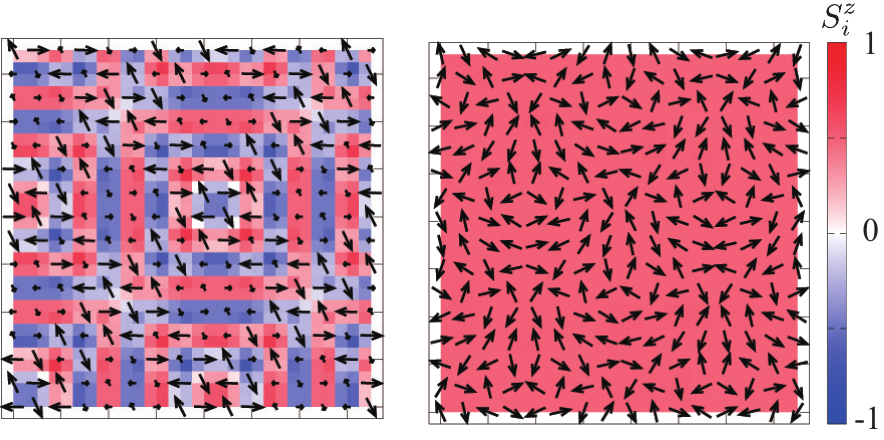} 
\caption{
\label{fig: CaseIII_spin_other} 
Real-space spin configurations of (left panel) the 4$Q$ I state for $H=0.1$ and (right panel) the 4$Q$ II state for $H=1$ at $K=0.02$ and $I^z=0.2$, which are obtained for different ordering wave vectors $\bm{Q}_1=(7\pi/8, \pi/8)$.  
The arrows represent the in-plane spin components, while the color represents the out-of-plane spin component. 
}
\end{center}
\end{figure}

Next, we turn on $I^z$ and $K$. 
Figure~\ref{fig: CaseIII_PD} shows the low-temperature phase diagram with changing $I^z$ and $H$ for small $K=0.02$. 
By introducing $K$, the 4$Q$ state becomes the ground state in the whole region in the phase diagram except for $H=2$, where the fully polarized state appears. 
There are two types of the 4$Q$ states, which are denoted as the 4$Q$ I state stabilized in the low-field region and the 4$Q$ II state stabilized in the high-field region. 

The 4$Q$ I state is characterized by a superposition of spin density waves at $\bm{Q}_1$--$\bm{Q}_4$ with equal intensity, as shown by the spin structure factor in Fig.~\ref{fig: CaseIII_spin}(c). 
The real-space spin configuration is shown in Fig.~\ref{fig: CaseIII_spin}(a), where a complicated noncoplanar spin texture happens, although there is no net scalar spin chirality in contrast to the skyrmion crystal. 
There are intensities at high-harmonic wave vectors, such as $-\bm{Q}_1+2\bm{Q}_4$, owing to a superposition of multiple spin density waves in both $xy$- and $z$-spin components, as shown in Fig.~\ref{fig: CaseIII_spin}(c). 
By increasing $H$, the 4$Q$ I state changes into the 4$Q$ II state with a jump of the magnetization. 
The spin configuration of the 4$Q$ II state corresponds to that in Eq.~(\ref{eq: 4QcaseIII}), where the real-space snapshot is presented in Fig.~\ref{fig: CaseIII_spin}(b) and the spin structure factor in momentum space is shown in Fig.~\ref{fig: CaseIII_spin}(d). 
This phase continuously changes into the fully polarized state at $H=2$. 

The qualitatively same phase diagram is obtained for another ordering wave vector, i.e., $\bm{Q}_1=(7\pi/8, \pi/8)$. 
We show the real-space spin configurations of the 4$Q$ I and 4$Q$ II states in the left and right panels of Fig.~\ref{fig: CaseIII_spin_other}, respectively. 
Compared to the spin configuration in Fig.~\ref{fig: CaseIII_spin}(a), the spin configuration of the 4$Q$ I state in the left panel of Fig.~\ref{fig: CaseIII_spin_other} seems to be different, although the spin structure factor shows similar profiles to each other. 
In this way, the instability toward the 4$Q$ state is expected when the ordering wave vectors belong to Case III.

\subsection{Case IV}
\label{sec: Case IV}

Finally, let us discuss the multiple-$Q$ instability in Case IV, where the ordering wave vector lies at $\bm{Q}_1=(3\pi/4, \pi/4)$, as denoted by the blue triangle in Fig.~\ref{fig: Qvec}. 
Since the magnetic phase diagram at this ordering wave vector has been already shown in previous literature~\cite{Hayami_PhysRevB.108.094415}, we briefly mention the difference from other cases. 

In Case IV, a checkerboard bubble lattice with the collinear 4$Q$ spin configuration appears in the low-field region, while the 4$Q$ conical state, whose spin configuration is similar to that in Eq.~(\ref{eq: 4QcaseIII}), appears in the high-field region when introducing a small biquadratic interaction $K$~\cite{Hayami_PhysRevB.108.094415}.
In other words, the instability toward the 4$Q$ state occurs at $\bm{Q}_1=(3\pi/4, \pi/4)$ similar to Case III. 
On the other hand, the spin configurations in the low-field region are different between Case III and Case IV; the noncoplanar spin configuration emerges in Case III [Fig.~\ref{fig: CaseIII_spin}(a)], while the collinear spin configuration emerges in Case IV.

\section{Summary}
\label{sec: Summary}

To summarize, we have investigated the multiple-$Q$ instability on the two-dimensional square lattice by focusing on the situation where the ordering wave vectors lie on the low-symmetric position in the Brillouin zone. 
By performing numerical simulations based on the simulated annealing for the spin model with various ordering wave vectors, we find that the multiple-$Q$ instabilities are classified into four cases. 
In Case I [Sec.~\ref{sec: Case I}], the ground state corresponds to the single-$Q$ state in the bilinear spin model. 
The anisotropic double-$Q$ state is induced by taking into account the effect of the biquadratic interaction ($K$). 
In Case II [Sec.~\ref{sec: Case II}], the single-$Q$ and double-$Q$ states are energetically degenerate in the bilinear spin model. 
The isotropic double-$Q$ state is chosen for small $K$. 
In Case III [Sec.~\ref{sec: Case III}], the single-$Q$, double-$Q$, and quadruple-$Q$ states are energetically degenerate in the bilinear spin model, and the quadruple-$Q$ state is chosen as the ground state when $K$ is considered. 
In Case IV [Sec.~\ref{sec: Case IV}], the situation is similar to that in Case III, although the quadruple-$Q$ spin configurations in the low-field region are different from each other. 
Our present results indicate the possibility of realizing further exotic multiple-$Q$ states according to the position of the ordering wave vectors, which will stimulate an experimental exploration in future studies.

\begin{acknowledgments}
This research was supported by JSPS KAKENHI Grants Numbers JP21H01037, JP22H04468, JP22H00101, JP22H01183, JP23H04869, JP23K03288, and by JST PRESTO (JPMJPR20L8) and JST CREST (JPMJCR23O4).  
Parts of the numerical calculations were performed in the supercomputing systems in ISSP, the University of Tokyo.
\end{acknowledgments}

\appendix

\bibliographystyle{apsrev}
\bibliography{../ref}
\end{document}